\documentclass[11pt,leqno]{article}

\usepackage[dvips]{epsfig}
\usepackage{color}
\usepackage{graphics}
\usepackage{amsmath}
\usepackage{amsthm}
\usepackage{amscd}
\usepackage{amssymb}

\newcommand{\ra}{\rangle}
\newcommand{\tens}{\otimes}

\title{A presentation of the quantum Fourier transform from a
  recursive viewpoint}
\author{Gloria \textsc{Paradisi}}
\date{Notes from a talk by Prof. Hugues \textsc{Randriam}\\ October 21, 2004}

\begin{document}

\maketitle

In most introductory texts to quantum computation, such
as \cite{ChNi} or \cite{Pres}, the presentation of the
quantum Fourier transform relies on the derivation of
an explicit formula, which is then translated into
a quantum circuit. The connection with the classical
fast Fourier transform is also cited, but not explicited,
or left as an exercise.

In this note we construct a quantum Fourier transform
circuit in a recursive way, by directly copying the construction
of the fast Fourier transform algorithm that is given,
for example, in \cite{CLRS}, chapter 30. We do not pretend
this presentation to be original, nor claim for any anteriority.
The aim of this paper is purely pedagogical.

The author would like to thank Prof. Hugues Randriam for his
supervision of the writing of this text.

\section{The fast Fourier transform}

Let $n$ be an integer and $N=2^n$. The discrete Fourier transform
associates to each complex $N$-tuple $a=(a_0,\dots,a_{N-1})$
the complex $N$-tuple $b=(b_0,\dots,b_{N-1})$ with
\begin{equation}
\label{defFourier}
b_k=2^{-n/2}\sum_{j=0}^{N-1}\zeta_n^{jk}a_j
\end{equation}
where
\begin{equation}
\zeta_n=e^{\frac{2i\pi}{2^n}}
\end{equation}
is a primitive $N$-th root
of $1$.

Thus, apart from some power of $\sqrt{2}$, the discrete Fourier transform
amounts to evaluating the polynomial
\begin{equation}
P(X)=\sum_{j=0}^{N-1}a_jX^j,
\end{equation}
of degree $2^n-1$, at the powers of $\zeta_n$.
The fast Fourier transform algorithm
performs this using a \emph{divide and conquer} strategy.
Obviously $P$ can be written in a unique way
\begin{equation}
\label{pair-impair}
P(X)=P_{even}(X^2)+XP_{odd}(X^2).
\end{equation}
In this decomposition the degree of $P_{even}$ (resp. $P_{odd}$)
is $2^{n-1}-1$, and its coefficients are the $a_j$ for
even (resp. odd) $j$'s. Using the fact that $\zeta_n^2=\zeta_{n-1}$,
one gets
\begin{equation}
\label{formulerec}
P(\zeta_n^k)=P_{even}(\zeta_{n-1}^k)+\zeta_n^kP_{odd}(\zeta_{n-1}^k).
\end{equation}
Thus, to perform a Fourier transform of order $n$, one has
to perform \emph{two} Fourier transforms of order $n-1$, and
then $O(n)$ additions and multiplications.
We will summarize this as
\begin{equation}
\label{FFTrec}
FFT_n=2.FFT_{n-1}+O(n)
\end{equation}
which finally gives us that the total number of elementary
operations required grows as
\begin{equation}
FFT_n=O(n2^n).
\end{equation}

\section{The quantum Fourier transform}

As we just saw, the exponential factor $2^n$ in the complexity
of the fast Fourier transform algorithm comes from the factor $2$
in equation \eqref{FFTrec}, that is, from the fact that the
computation of a Fourier transform of order $n$ requires \emph{two}
Fourier transforms of order $n-1$. 
Hopefully this factor $2$ could be shrunk down to $1$ using
quantum parallelism. Indeed, recall that when one has a quantum
circuit $F$ acting on $n-1$ qubits, then the circuit $F\tens\mathbf{1}$
(see figure \ref{Ftens1} below)
\begin{figure}[ht]
\begin{center}
\input{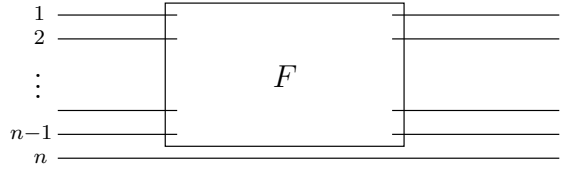}
\caption{constructing $F\tens\mathbf{1}$ out of $F$}\label{Ftens1}
\end{center}
\end{figure}
on $n$ qubits, applied to a vector $|x\ra=|x_0\ra|0\ra+|x_1\ra|1\ra$,
gives as output $(F|x_0\ra)|0\ra+(F|x_1\ra)|1\ra$,
thus, in some way, computing
simultaneously $F|x_0\ra$ and $F|x_1\ra$ at the cost of only
\emph{one} application of $F$.
Carrying this out succesfully
will lead to a dramatic decrease of complexity in the adaptation
of the Fourier transform from the classical world to the quantum world:
from an exponential complexity we will get to a quadratic one.

To make it formally, consider an $n$ qubits system, and denote
the vectors of the computational basis by
\begin{equation}
|j\ra=|j_1\ra\dots|j_{n-1}\ra|j_n\ra,
\end{equation} 
for $j\in\{0,\dots,N-1\}$ (recall that $N=2^n$) with base $2$ expansion
\begin{equation}
j=j_12^{n-1}+\dots+j_{n-1}2+j_n.
\end{equation}
We will construct a quantum circuit which sends the unitary
vector $\sum_ja_j|j\ra$ to $\sum_kb_k|k\ra$
where the $b_k$'s are deduced from the $a_j$'s by relation 
\eqref{defFourier}.
Remark that, since quantum circuits can only perform unitary
transformations,
this implies that formula \eqref{defFourier} is norm preserving.
Although this fact is well known (Parseval's theorem), we will
not need to assume it. Indeed our construction
can be considered as a way of re-proving it.

Now let
\begin{equation}
|P\ra=\sum_{j=0}^{N-1}a_j|j\ra.
\end{equation}
Grouping together terms according to $j_n=0$ (that is, $j$ even)
or $j_n=1$ (that is, $j$ odd) we get a decomposition
\begin{equation}
|P\ra=|P_{even}\ra|0\ra+|P_{odd}\ra|1\ra,
\end{equation}
which is the exact analogue of decomposition \eqref{pair-impair}.

According to formula \eqref{formulerec}, we have then to perform
a Fourier transform on the first $n-1$ qubits, which can be done
as in figure \ref{Ftens1}. In doing that one has however to be
careful on the ordering of the output qubits. Indeed, write
\begin{equation}
k=k_12^{n-1}+k_22^{n-2}+\dots+k_{n-1}2+k_n.
\end{equation}
Then, as $\zeta_{n-1}^{2^{n-1}}=1$
and $\zeta_n^{2^{n-1}}=-1$, formula \eqref{formulerec}
can be re-writen as
\begin{equation}
\label{formulerecbis}
P(\zeta_n^k)=P_{even}(\zeta_{n-1}^{k'})+(-1)^{k_1}\zeta_n^{k'}P_{odd}(\zeta_{n-1}^{k'})
\end{equation}
where
\begin{equation}
k'=k_22^{n-2}+\dots+k_{n-1}2+k_n.
\end{equation}
Thus, while $|P_{even}\ra$ and $|P_{odd}\ra$ are vector states of
the \emph{first} $n-1$ qubits $j_1,\dots,j_{n-1}$,
their Fourier transforms should
be output on the \emph{last} $n-1$ qubits $k_2,\dots,k_n$.
To circumvent this difficulty we will consider a \emph{modified}
quantum Fourier transform, $mQFT$,
which reverses the order of the output qubits.
Obviously this is not a serious problem, since the correct order can
then be restablished using only $n/2$ swap gates.

Our modified quantum Fourier transform circuit of order $n$ should
then look like:
\begin{center}
\begin{figure}[h]
\begin{center}
\input{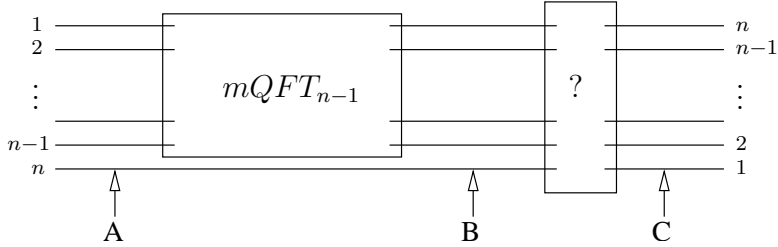}
\caption{recursive scheme for $mQFT_n$.}\label{schema}
\end{center}
\end{figure}
\end{center}

At point A we input the vector
\begin{equation}
|P\ra=|P_{even}\ra|0\ra+|P_{odd}\ra|1\ra,
\end{equation}
so that at point $B$ we get
\begin{equation}
\label{B}
|\widetilde{P_{even}}\ra|0\ra+|\widetilde{P_{odd}}\ra|1\ra,
\end{equation}
where we denoted by $\widetilde{\;.\;}$ the effect of our (modified)
quantum Fourier transform.
According to formula \eqref{formulerecbis}
we have
\begin{equation}
P(\zeta_n^k)=
\begin{cases}
P_{even}(\zeta_{n-1}^{k'})+\zeta_n^{k'}P_{odd}(\zeta_{n-1}^{k'}) &
\text{for $k_1=0$}\\
P_{even}(\zeta_{n-1}^{k'})-\zeta_n^{k'}P_{odd}(\zeta_{n-1}^{k'}) &
\text{for $k_1=1$}
\end{cases}
\end{equation}
so that, putting the $2^{-n/2}$ back,
at point C we should get
\begin{equation}
\label{C}
|\widetilde{P}\ra=\frac{1}{\sqrt{2}}(|\widetilde{P_{even}}\ra+\zeta_n^{k'}|\widetilde{P_{odd}}\ra)|0\ra+\frac{1}{\sqrt{2}}(|\widetilde{P_{even}}\ra-\zeta_n^{k'}|\widetilde{P_{odd}}\ra)|1\ra,
\end{equation}
where $\zeta_n^{k'}$ can be understood as the diagonal matrix acting
on the first
$n-1$ qubits by multiplying $|k'\ra$ by $\zeta_n^{k'}$, or
equivalently as the controlled operation that multiplies the last
qubit by $\zeta_n^{k'}$ when the first $n-1$ qubits are set to $|k'\ra$.

Clearly one can go from \eqref{B} to \eqref{C} by sending $|0\ra$
to $\frac{1}{\sqrt{2}}(|0\ra+|1\ra)$ and $|1\ra$
to $\frac{1}{\sqrt{2}}(\zeta_n^{k'}|0\ra-\zeta_n^{k'}|1\ra)$,
that is by acting on the last qubit through
\begin{equation}
\frac{1}{\sqrt{2}}\begin{bmatrix}1 & \zeta_n^{k'} \\ 1 &
  -\zeta_n^{k'}\end{bmatrix}=HR_2^{k_2}\dots R_{n-1}^{k_{n-1}}R_n^{k_n},
\end{equation}
where $H=\frac{1}{\sqrt{2}}\begin{bmatrix}1 & 1 \\ 1 &
  -1\end{bmatrix}$
is the usual Hadamard matrix and $R_d=\begin{bmatrix}1 & 0 \\ 0 &
  e^{\frac{2i\pi}{2^d}}\end{bmatrix}$.

This means that we can complete our quantum Fourier circuit by
replacing the '?' circuit in figure \ref{schema} with
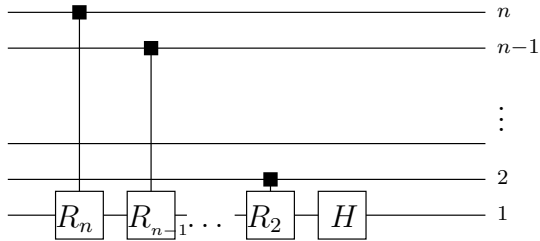
\begin{center}
\begin{figure}[h]
\begin{center}
\input{final.pstex_t}
\end{center}
\end{figure}
\end{center}

Thus we see that the (modified) quantum Fourier circuit of order $n$
can be constructed from the order $n-1$ one by adding $n-1$ controlled
operations and one Hadamard gate. Putting all together, since the
quantum Fourier circuit of order $0$ is trivial (the identity), we
get a total complexity of $\frac{n(n+1)}{2}$ gates.

The reader will remark that the circuit we get is the inverse of
the one that is given in \cite{ChNi}, figure 5.1. This is not a
mistake: it only indicates that the Fourier transform is
essentially its own inverse.

\bigskip

\begin{flushleft}
Gloria \textsc{Paradisi}, c/o Hugues \textsc{Randriam}\\
Ecole nationale sup\'erieure des t\'el\'ecommunications\\
46, rue Barrault\\
75634 Paris Cedex 13\\
France\\
\medskip
\texttt{Gloria.Paradisi@salle-s.org}
\end{flushleft}

\end{document}